# High-field Breakdown and Thermal Characterization of Indium Tin Oxide Transistors

Haotian Su[1,†], Yuan-Mau Lee[2,†], Tara Peña[1], Sydney Fultz-Waters[2], Jimin Kang[1], Çağıl Köroğlu[1], Sumaiya Wahid[1], Christina J. Newcomb[3], Young Suh Song[1], H.-S. Philip Wong[1], Shan X. Wang[1,2], and Eric Pop[1,2,4,5,*]

[1]*Department of Electrical Engineering, Stanford University, Stanford, CA 94305, USA*

[2]*Department of Materials Science and Engineering, Stanford University, Stanford, CA 94305, USA*

[3]*Stanford Nano Shared Facilities, Stanford University, Stanford, CA 94305, USA*

[4]*Department of Applied Physics, Stanford University, Stanford, CA 94305, USA*

[5]*Precourt Institute for Energy, Stanford University, CA 94305, USA*

[†]*Authors contributed equally.* [*]*Contact:* epop@stanford.edu

**ABSTRACT:** Amorphous oxide semiconductors are gaining interest for logic and memory transistors compatible with low-temperature fabrication. However, their low thermal conductivity and heterogeneous interfaces suggest that their performance may be severely limited by self-heating, especially at higher power and device densities. Here, we investigate the high-field breakdown of ultrathin (~4 nm) amorphous indium tin oxide (ITO) transistors with scanning thermal microscopy (SThM) and multiphysics simulations. The ITO devices break irreversibly at channel temperatures of ~180 °C and ~340 °C on $SiO_2$ and $HfO_2$ substrates, respectively, with failure primarily caused by thermally-induced compressive strain near the device contacts. Combining SThM measurements with simulations allows us to estimate a thermal boundary conductance (TBC) of 35 ± 12 $MWm^{-2}K^{-1}$ for ITO on $SiO_2$ and 51 ± 14 $MWm^{-2}K^{-1}$ for ITO on $HfO_2$. The latter also enables significantly higher breakdown power due to better heat dissipation and closer thermal expansion matching. These findings provide insights into the thermo-mechanical limitations of indium-based amorphous oxide transistors, which are important for more reliable and high-performance logic and memory applications.

## INTRODUCTION

Amorphous oxide semiconductors (AOS) are well-established in the display industry[1-3] and are increasingly recognized as promising back-end-of-line (BEOL)-compatible channel materials for thin-film transistors.[4-6] Among them, indium tin oxide (ITO) transistors stand out due to their low-temperature large-scale deposition methods, high drive current, and low leakage, making them promising for *n*-type BEOL logic and memory applications.[5-10] However, their performance and stability could be limited by self-heating during operation,[10-15] with heat dissipation challenges potentially worsened by high power densities and high device densities.[16-19]

Broadly speaking, thermal management is a critical challenge across all modern electronics, impacting not only transistor performance[20-22] but also memory,[23,24] displays,[25-27] and integrated circuit reliability,[28-30] as well as flexible electronics[31,32] where self-heating is amplified by the low thermal conductivity of the substrates. To address these thermal challenges in oxide transistors, various strategies have been explored. P.-Y. Liao *et al*.[13] incorporated an interlayer between the transistor channel and substrate to facilitate device heat dissipation, C. Besleaga *et al*.[33] employed a gate dielectric with high thermal conductivity to reduce self-heating, while K. Kise *et al*.[34] implemented a U-shaped transistor design to reduce self-heating. Furthermore, high thermal conductivity substrates, such as high-resistivity Si, SiC, and diamond, have also been explored to aid heat dissipation and alleviate self-heating effects.[10,12,14,35]

Despite these advancements, the heat dissipation and breakdown mechanisms in ITO transistors remain poorly understood. Moreover, the thermal boundary conductance (TBC) at the interface between the ITO channel and its dielectric, essential for transistor heat dissipation,[32,36-38] is underexplored. Techniques like thermoreflectance imaging and time-domain thermoreflectance have been widely used to study the thermal properties of devices and thin films.[12,35,39-41] However, their limitations, such as low spatial resolution (i.e. spot size greater than transistor dimensions) and considerable uncertainties in TBC measurements for interfaces between ultrathin amorphous films, highlight the need for alternative approaches to understand the thermal properties and breakdown mechanisms of ITO transistors.

In this study, we fabricate ITO transistors with 4 nm thin sputtered channels and investigate their breakdown during high-field operation. Using scanning thermal microscopy (SThM),[42,43] we measure the ITO channel temperature during operation and, through simulations, quantify the TBC at ITO-$SiO_2$ and ITO-$HfO_2$ interfaces. These interfaces limit heat dissipation and, along with their thermal expansion mismatch, contribute to localized compressive strain. Our analysis reveals that ITO transistor breakdown is driven by thermally-accelerated cracks that appear near the ITO channel/contact edges. These findings provide insights into the thermo-mechanical limitations of ITO transistors and inform the design of more reliable future devices, with implications for the broader nanotechnology community working on thermal management and reliability in nanoscale devices.

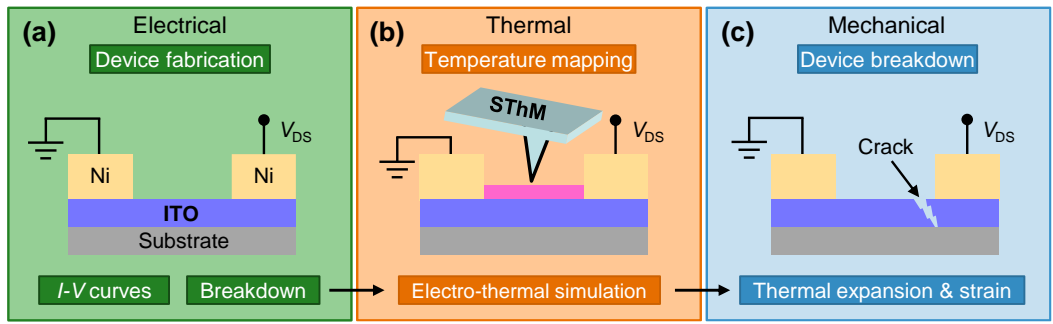

**Figure 1**. Overview of our electro-thermo-mechanical multiphysics approach. Electrical analysis begins with **(a)** ITO transistor fabrication, measuring current-voltage up to device breakdown. Thermal analysis in **(b)** is done with scanning thermal microscopy (SThM), combined with finite-element electro-thermal modeling, to extract thermal properties such as the thermal boundary conductance (TBC) of the ITO-$SiO_2$ interface. Solid mechanics simulations, **(c)**, reveal thermal expansion and peak strain distributions which appear consistent with images of cracks forming in the ITO channel, leading to device failure.

## RESULTS AND DISCUSSION

We employ an electro-thermo-mechanical multiphysics approach to investigate the failure mechanisms and heat dissipation in ITO transistors, as depicted in **Figure 1**. Electrical analysis begins with the fabrication of back-gated ITO transistors, using a 4 nm thick sputtered ITO channel on a SiO$_2$ (100 nm) on p$^{++}$ Si substrate with Ni top contacts. The highly-doped Si substrate is also used as the back-gate and more fabrication details can be found in S. Wahid *et al.*,[9] as well as in our **Methods section** and Supporting Information **Section S1**. Nearly one hundred devices were characterized electrically, up to their breakdown, with channel lengths (*L*) between 1.5 to 1.7 μm and widths of 10 μm.

To map the device temperature during operation we used SThM,[37,44-46] which has sub-100 nm spatial resolution, depending on the probing tip and environmental conditions. (Raman thermometry[37,45,46] appears impractical with the ultrathin amorphous ITO, which does not have a usable Raman signal.) These thermal measurements were complemented with finite-element modeling,[23,32,37] which enabled simulation of device temperatures and estimation of some unknown parameters, such as TBC. Simulations of mechanical strain distributions across the transistor[47,48] were also conducted to assess the impact of thermal expansion during operation. Comparing these simulations with experimental observations of device failure correlates electrical performance, heat dissipation, and mechanical reliability in our ITO devices.

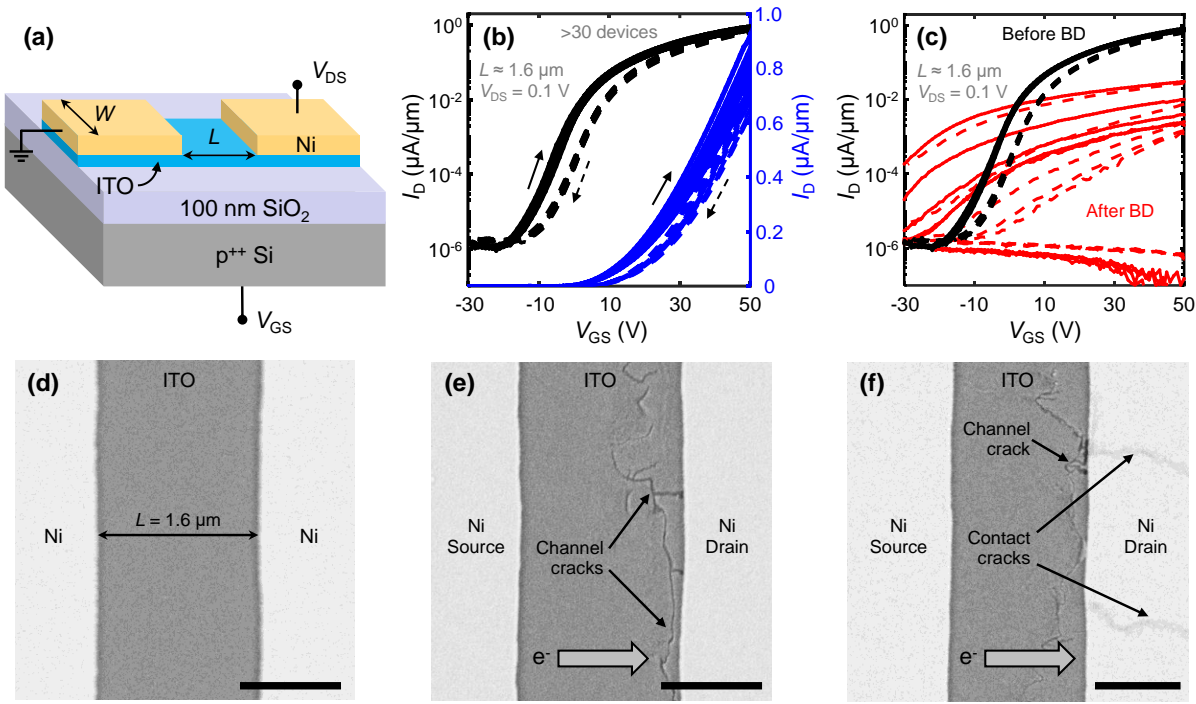

**Figure 2**. **(a)** Schematic of back-gated ITO transistors, with 4 nm thin ITO channel on SiO$_2$ (100 nm)/Si (p$^{++}$) substrate, and 80 nm Ni source and drain contacts. **(b)** Measured transfer curves from > 30 devices with channel length $L \approx 1.6$ μm, on log (black) and linear scale (blue). **(c)** Transfer curves for a subset of 10 devices before (black) and after (red) breakdown (BD), showing significant reduction of on-state $I_D$ and loss of gate control. Solid/dashed lines mark forward/backward $V_{GS}$ sweeps. **(d)** Scanning electron microscopy (SEM) image of initial ITO transistor channel compared with **(e)** SEM image of a device after breakdown showing channel cracks, and **(f)** another device showing additional cracks under the Ni drain. Scale bars in (d-f) are all 1 μm. Block arrows in (e, f) show direction of electron flow.

**Electrical Characterization**

ITO transistors with the geometry illustrated in **Figure 2a** were fabricated and characterized to evaluate their electrical performance and breakdown. Transfer curves ($I_D$ vs. $V_{GS}$) shown in **Figure**

**2b** were measured for > 30 devices with channel length of ~1.6 μm, revealing relatively low variability. Breakdown measurements were conducted on a subset of 10 devices by applying $V_{GS}$ = 45 V and sweeping $V_{DS}$ from 0 V to 40 V. Transfer curves measured after breakdown, in **Figure 2c**, show much lower on-state current and loss of gate control. Electrical characterization details are provided in the **Methods section**. Scanning electron microscopy (SEM) images further reveal the physical damage in these transistors: **Figure 2d** compares an original device channel with failed devices in **Figure 2e** (prominent cracks in the ITO channel near the drain) and **Figure 2f** (additional cracks in the Ni drain contact).

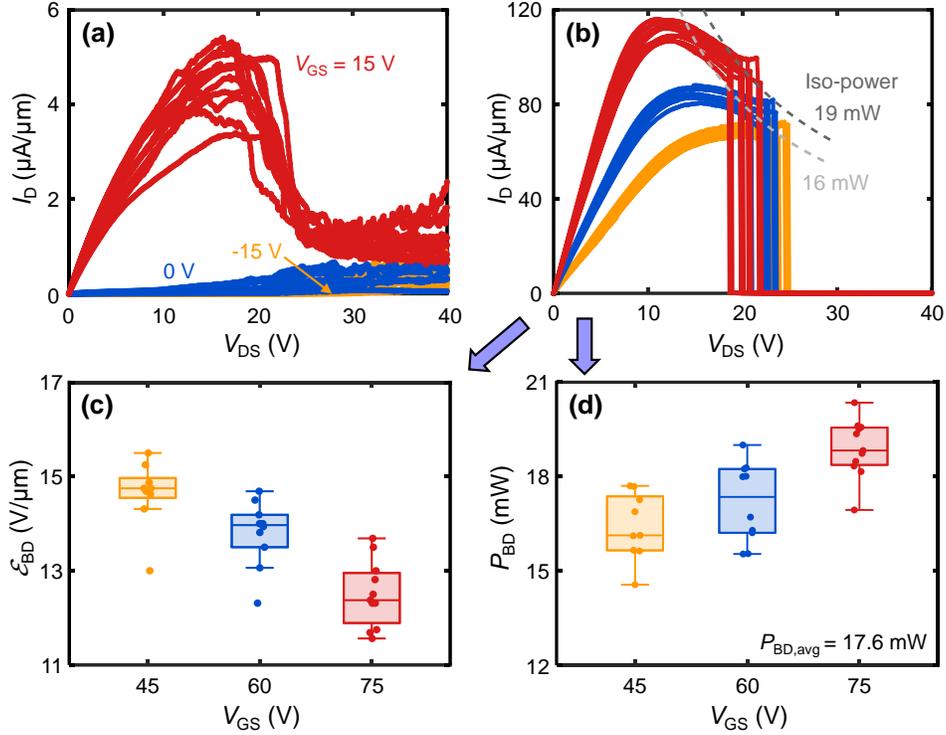

**Figure 3.** (a) Measured current vs. drain voltage ($I_D$ vs $V_{DS}$) curves of ITO transistors in sub-threshold or low overdrive ($V_{GS}$ = -15, 0, 15 V). (b) $I_D$ vs $V_{DS}$ breakdown of ITO transistors at high $V_{GS}$ = 45, 60, 75 V, showing an abrupt $I_D$ drop around $V_{DS} \approx$ 20 V; ~10 devices were measured for each $V_{GS}$. Dashed lines mark the range of power dissipation at device breakdown. (c) Average lateral electric field at device breakdown ($\mathcal{E}_{BD}$) along the ITO channel, from the three $V_{GS}$ conditions in (b). (d) Breakdown power ($P_{BD}$) under the same breakdown $V_{GS}$ in (b), with average breakdown power $P_{BD,avg} \approx$ 17.6 mW. We note that all electrical measurements in this work are performed in direct current (DC) mode, to simplify the thermal analysis and keep them consistent with the SThM measurements. Pulsed measurements[49-51] could, in principle, be used to study operation with reduced self-heating, but they are impractical here because the electrical time constant of our devices, which is dominated by the large pad capacitance, is much greater than their thermal time constant, which is expected to be around ~30 ns, dominated by the SiO$_2$ substrate.[52]

Current vs. drain voltage ($V_{DS}$) measurements up to device breakdown are shown in **Figures 3a-b**, with different gate voltages ($V_{GS}$); in **Figure 3a**, devices are operated below threshold ($V_{GS}$ = -15, 0 V) and at smaller overdrive ($V_{GS}$ = 15 V). Below threshold, $I_D$ increases weakly with $V_{DS}$ due to carrier extraction from trap states under increasing lateral electric field, consistent with previous reports for AOS-based power devices.[53-55] However, for device breakdowns in the on-state (**Figure 3b** with $V_{GS}$ = 45, 60, 75 V), an abrupt current drop is observed at $V_{DS} \approx$ 20 V, similar to breakdown behavior recently reported in ~10 times larger and thicker IGZO devices.[56] Additional details on transfer curves and gate leakage before and after breakdown are provided in Supporting Information **Section S2**. The average electric field at breakdown ($\mathcal{E}_{BD}$) along the ITO channel decreases with increasing $V_{GS}$ (**Figure 3c**), suggesting that the breakdown is initiated by higher current and higher temperature (i.e. higher carrier density at higher $V_{GS}$) rather than the lateral electric field acting alone. The corresponding breakdown power ($P_{BD}$) in **Figure 3d** displays a weak increase with $V_{GS}$, which is consistent with improved field uniformity as the transistor breaks down deeper in the linear regime at higher $V_{GS}$. Nevertheless, as we will later see, the ITO channel breakdown mechanism is more complex.

**Thermal Characterization**

To evaluate the device temperature during operation, we employ scanning thermal microscopy (SThM). The experimental setup, shown schematically in **Figure 4a**, utilizes a Wheatstone bridge to connect the SThM cantilever, which operates in contact mode. Devices were capped with a 6 nm $Al_2O_3$ layer to prevent direct electrical contact[37,44,45] between the SThM probe and ITO channel. More details on the SThM setup and calibration are provided in the **Methods section** and Supporting Information **Section S3**. The transfer curves of uncapped and $Al_2O_3$-capped devices (**Figure 4b**) reveal a negative shift in threshold voltage for the capped devices, consistent with our prior findings.[9] This capping layer has negligible impact on the ITO channel temperature measured by SThM.[37] For SThM measurements, to prevent device damage or breakdown, the devices were measured in the linear $I_D$ vs. $V_{DS}$ region, a subset of which is shown in **Figure 4c**.

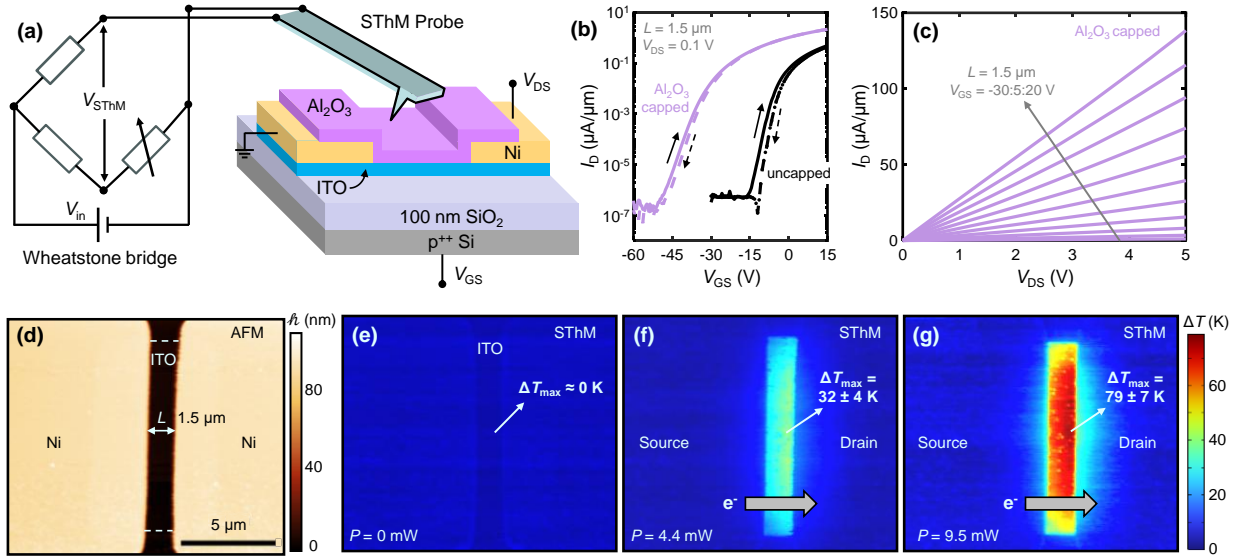

**Figure 4.** (a) Schematic of SThM measurement on ITO transistors, showing a Wheatstone bridge connected to the SThM cantilever. The tip scans in contact mode with the top device surface. (b) Measured transfer curves of uncapped (black) and $Al_2O_3$-capped (purple) devices, showing negative shift in threshold voltage consistent with our previous work.[9] Solid/dashed lines mark forward/backward $V_{GS}$ sweeps. (c) Measured $I_D$ vs. $V_{DS}$ curve of $Al_2O_3$-capped devices, showing linear behavior. (d) AFM topography scan of an ITO transistor, and corresponding SThM temperature map at input powers of: (e) 0 mW, (f) 4.4 mW, and (g) 9.5 mW. The block arrow shows the direction of electron flow, from source to drain. At an input power of 9.5 mW, the maximum ITO temperature rise ($\Delta T_{max}$, above room temperature) is $79 \pm 7$ K, as shown in (g).

The atomic force microscopy (AFM) topography scan of an ITO transistor with 1.5 μm long channel is shown in **Figure 4d**. The corresponding SThM maps (**Figures 4e–g**) display temperature distributions at different input powers, $P = I_D V_{DS}$. **Figure 4e**, at 0 mW, displays the background temperature rise ~0 K, as expected. At 4.4 mW and 9.5 mW input powers, $\Delta T_{max}$ reaches 32 ± 4 K and 79 ± 7 K, respectively. The uncertainties arise from SThM calibration, with more details in Supporting Information **Section S3**. The measured temperature is slightly higher near the drain contact (**Figure 4g**), which is expected due to the direction of electron flow.[37] (The electron density is lower and the lateral field is higher near the drain.) The temperature drops rapidly at the Ni contacts, which function as heat sinks.

**Electro-Thermo-Mechanical Modeling**

We also carried out finite-element simulations[23,32,37] to complement the SThM measurements and understand the heat dissipation characteristics of ITO transistors, with modeling details in **the Methods section.** The simulated temperature map with 9.5 mW input power is shown in **Figure 5a**, revealing a peak temperature rise $\Delta T_{max} = 71.7$ K. This simulation uses thermal conductivity ($k$) and thermal boundary conductance (TBC) values detailed in Supporting Information **Section S4**, and it assumes uniform power dissipation across the device channel, which is a good approximation for a device operating in the linear region.[57] This approximation does not capture the subtle temperature rise near the drain seen experimentally (**Figure 4g**), but is sufficiently good to yield an *average* estimate of the TBC (between ITO and its substrate), which dominates heat flow. We also performed a sensitivity analysis with respect to some of the key thermal simulation parameters – as shown in **Figure 5b**, $\Delta T_{max}$ has negligible dependence on practical $k$ ranges for $Al_2O_3$, Ni, and ITO. Similarly, **Figure 5c** indicates that wide TBC variations (25–250 MWm$^{-2}$K$^{-1}$) for ITO-Ni, ITO-$Al_2O_3$, and Ni-$Al_2O_3$ interfaces have minimal impact on $\Delta T_{max}$, suggesting those interfaces play a minimal role in heat sinking. However, $\Delta T_{max}$ is strongly dependent on the TBC of the ITO-$SiO_2$ interface, which is estimated to be 35 ± 12 MWm$^{-2}$K$^{-1}$, by comparing our simulations with SThM measurements of the peak temperature (79 ± 7 K in **Figure 4g**) for this power input. The TBC$_{ITO-SiO2}$ estimated here is consistent with other TBCs reported for similar interfaces.[13,58]

**Figure 5d** shows the relationship between $\Delta T_{max}$ and input power ($P$), demonstrating good agreement between temperatures measured by SThM and simulated temperature at the average breakdown power ($P_{BD,avg}$, determined in **Figure 3d**). The average breakdown temperature ($T_{BD}$) of our ITO transistors on 100 nm $SiO_2$/Si substrate is found to be ~155-181 °C. This temperature is lower than the annealing temperature of ITO transistors,[8,9] which can reach up to 300 °C. Therefore, we cannot attribute the breakdown of our ITO transistors solely to the temperature rise during operation, and other mechanisms must be at play. To understand other effects, we also carried out thermo-mechanical simulations, displaying the estimated strain distribution along the ITO channel, $\varepsilon_{xx}$, before current flow (black line) and after $P_{BD,avg}$ is applied (red line), as shown in **Figure 5e**. Initially, tensile strain is observed near the channel-contact edges and compressive strain appears under the contacts, which is attributed to the presence of Ni contacts.[48] (We report stress from a 80 nm evaporated Ni layer in Supporting Information **Section S5**.) With $P_{BD,avg}$ input power the device self-heats and the mismatch in coefficient of thermal expansion (CTE, $\alpha$) between ITO ($\alpha_{ITO} \approx 8 \times 10^{-6}$ K$^{-1}$)[59,60] and $SiO_2$ ($\alpha_{SiO2} \approx 5.6 \times 10^{-7}$ K$^{-1}$)[61,62] generates compressive strain at the channel edges (red lines in **Figure 5e**). The position of these compressive strain peaks corresponds to the locations of channel cracks seen in the SEM images from **Figures 2e-f**.

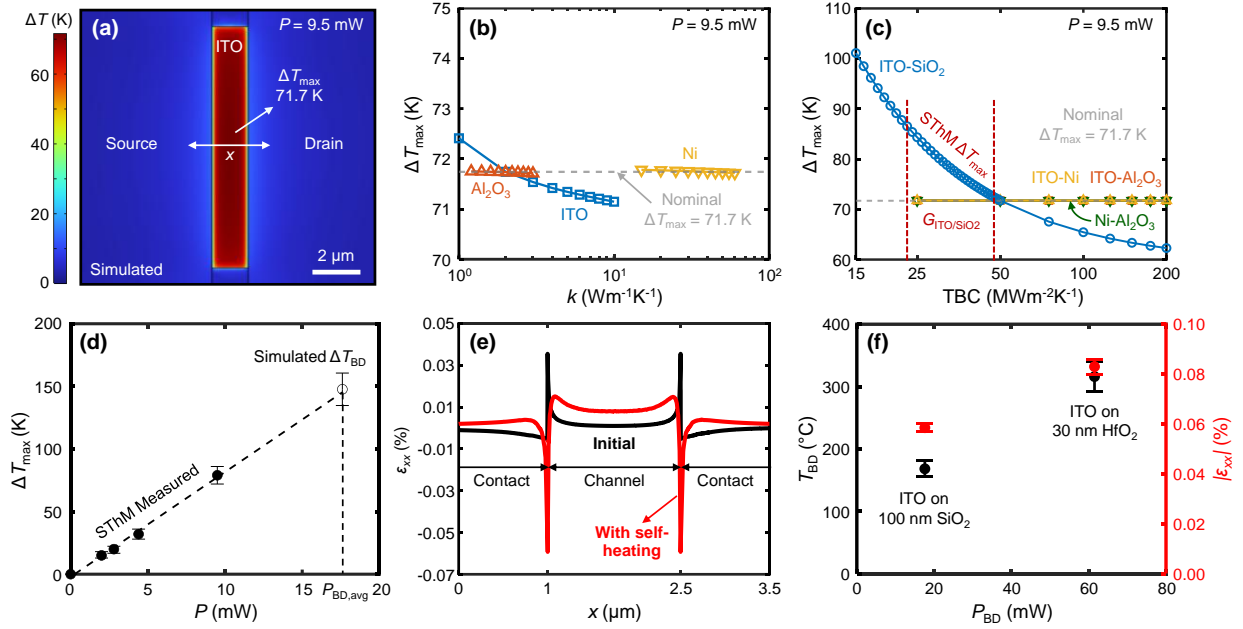

**Figure 5.** (a) Simulated temperature map of ITO transistor with 9.5 mW input power, showing peak temperature rise $\Delta T_{max} = 71.7$ K, with nominal thermal parameters provided in Supporting Information **Section S4**. (b) Sensitivity analysis of $\Delta T_{max}$ for the same device, with respect to the expected thermal conductivity ($k$) range of $Al_2O_3$, Ni, and ITO. (c) Sensitivity analysis of $\Delta T_{max}$ with respect to TBC at material interfaces, showing minimal dependence on TBC at ITO-Ni, ITO-$Al_2O_3$, and Ni-$Al_2O_3$ interfaces. However, $\Delta T_{max}$ varies with the TBC of ITO-$SiO_2$ interface, estimated to be $35 \pm 12$ MWm$^{-2}$K$^{-1}$ by comparing to the SThM measurements. (d) Dependence of $\Delta T_{max}$ on input power, $P$. Filled symbols mark SThM-measured temperature, hollow symbol is a simulation of $\Delta T_{max}$ at $P_{BD,avg}$ from in **Figure 3d**. Dashed line is a linear fit, to highlight the trend. (e) Initial strain distribution ($\varepsilon_{xx}$) along the '$x$' white arrow in panel (a), showing tensile strain in the ITO channel and compressive strain under contacts (black curve). At device breakdown with self-heating, the compressive strain peaks at the channel/contact edge (red curve). (f) Summary of breakdown temperature ($T_{BD}$) and peak compressive strain ($|\varepsilon_{xx}|$) vs. breakdown power for ITO transistors on 100 nm $SiO_2$ and on 30 nm $HfO_2$ dielectric, on Si back-gate.

## Discussion and Comparison to ITO on $HfO_2$

We note that the ~0.06% compressive strain suggested by our device simulations is lower than typical crack onset strain (COS) values previously reported for ITO, which range from 0.1% to several percent.[63-67] Part of this may be due to non-uniformities in our devices and contacts (e.g. Ni contact grains or non-uniform current flow at high input power), which cannot be captured by simulations that assume uniform material properties. In addition, the COS is also expected to depend on ITO thickness, substrate, and deposition conditions, and most reported COS values were under externally applied mechanical stresses. In contrast, cracking of ultrathin ITO due to compressive strain induced by electrical self-heating has not been previously explored.

Our findings suggest that the cracking failure of these ultrathin ITO transistors on $SiO_2$ is caused by the CTE mismatch between ITO and $SiO_2$, and initiated by self-heating effects during device operation. This CTE mismatch generates compressive strain near the channel edges, which appears to exceed the reduced COS of the ITO material at the elevated temperature. We also note that the low crystallization temperature of ITO (~150-200 °C)[63,64,68-71] may accelerate structural changes during device self-heating, such as grain boundary formation and localized (e.g. filamentary) crystallization.[10] A previous study[64] also suggested that defects and grain boundaries, often introduced during low-temperature annealing, can act as stress concentrators, accelerating crack formation. These microstructural changes, coupled with thermal and electrical stresses, likely create a cascading effect that weakens the ITO channel and promotes crack formation. While our SEM images and finite-element modeling provide insights into crack formation, transmission electron microscopy (TEM) analysis could further elucidate the crack propagation mechanisms for nanoscale devices and is recommended for future studies.

For comparison, we also fabricated sputtered ITO transistors on 30 nm $HfO_2$ on Si ($p^{++}$) substrates, with more details provided in Supporting Information **Section S6**. Electrical and SThM characterizations, along with finite-element simulations, were performed following the same methodology as described in **Figure 4** and **Figure 5**. The thermal boundary conductance (TBC) at the ITO-$HfO_2$ interface was $51 \pm 14$ MWm$^{-2}$K$^{-1}$, a value not previously available in the literature. While this TBC$_{ITO-HfO2}$ value is nearly 50% higher than the TBC$_{ITO-SiO2}$, it still falls on the lower end of typical TBC values for material interfaces.[20,72-74] SThM measurements and simulations (Supporting Information **Section S6**) further revealed that the breakdown temperature of ITO transistors on 30 nm $HfO_2$ lies between 292 °C and 340 °C, nearly double the $T_{BD}$ of ITO transistors on 100 nm $SiO_2$ substrate (both on Si). Despite the higher $T_{BD}$, the maximum compressive strain at device breakdown was estimated to be only ~0.086%, attributed to the closer CTE matching between ITO ($\alpha_{ITO} \approx 8 \times 10^{-6}$ K$^{-1}$) and $HfO_2$ ($\alpha_{HfO2} \approx 6 \times 10^{-6}$ K$^{-1}$).[75,76] **Figure 5f** highlights these differences by summarizing the $T_{BD}$ and peak compressive strain ($|\varepsilon_{xx}|$) at the breakdown power for ITO transistors on both substrates.

These findings suggest that sputtered ITO transistors on 30 nm $HfO_2$ have better heat dissipation and reduced thermal stress, enabling them to sustain higher power before breakdown. These results highlight the importance of efficient heat dissipation and CTE matching[77] in optimizing material systems for high-performance ITO transistors. Finally, as an additional comparison, we have also examined several ITO transistors prepared by atomic layer deposition (ALD), and observed similar breakdown behavior. Further details can be found in Supporting Information **Section S7**, where we also discuss how our findings could be extended to other amorphous oxide semiconductors.

**CONCLUSION**

We investigated the high-field breakdown of ITO transistors using an electro-thermo-mechanical multiphysics approach. Comparing scanning thermal microscopy measurements with simulations, we obtained the steady-state device temperature and estimated the thermal boundary conductance (TBC) at the ITO-$SiO_2$ interface, $35 \pm 12$ MWm$^{-2}$K$^{-1}$. This relatively low TBC, combined with the significant mismatch in coefficient of thermal expansion (CTE) between ITO and $SiO_2$, induces compressive strain near the contacts during device operation and leads to breakdown. It is also possible that the low crystallization temperature of ITO amplifies microstructural changes, such as grain boundary formation, and further accelerates failure under thermal and electrical stress. For comparison, we also fabricated ITO transistors on 30 nm $HfO_2$, on the same Si back-gate substrates. The ITO-$HfO_2$ TBC is found to be $51 \pm 14$ MWm$^{-2}$K$^{-1}$, approximately 50% higher than for ITO-$SiO_2$, which, combined with the thinner dielectric and closer CTE matching to ITO, enabled higher device breakdown power and temperature. This study provides insights into the electro-thermo-mechanical behavior of indium-based amorphous oxide transistors, underscoring the importance of heat dissipation and thermal stress management for their applications.

**METHODS**

**Device Fabrication.** The fabrication began with either a thermally-grown 100 nm $SiO_2$ layer or a plasma-enhanced atomic layer deposited (PEALD) 30 nm $HfO_2$ layer, both on $p^{++}$ Si substrates, which also serve as back-gates. Next, the 4 nm amorphous indium tin oxide (ITO) channel was deposited at room temperature using AJA magnetron sputtering, under a base pressure of $2 \times 10^{-8}$ Torr or lower. ITO was RF-sputtered at 100 W in a 5 mTorr argon-oxygen (5:1) atmosphere, with a deposition rate of 9.6 Å/min. The ITO channel was then patterned by optical lithography (Heidelberg MLA150) and wet etching in a 1.7% HCl solution. Finally, nickel (Ni) contacts were deposited using a Kurt J. Lesker electron-beam evaporator, with a base pressure of $\sim 5 \times 10^{-8}$ Torr and a deposition rate of 1 Å/s, patterned by lift-off.

**Electrical Characterization.** Electrical measurements were performed using a Keithley 4200 parameter analyzer with a Cascade Summit probe station, in air and room temperature ambient. Transfer curves ($I_D$ vs. $V_{GS}$) were measured for over 90 devices by sweeping $V_{GS}$ while keeping $V_{DS} = 0.1$ V. Breakdown measurements were conducted on a subset of 10 devices for each $V_{GS}$, sweeping $V_{DS}$ from 0 V to 40 V. Scanning electron microscopy (SEM) images were taken with a Thermo Fisher Apreo SEM to examine devices after breakdown. Device transfer curves after breakdown were also measured.

**Thermal Characterization.** Prior to scanning thermal microscopy (SThM) measurements, devices were capped with a 6 nm $Al_2O_3$ layer deposited via plasma-enhanced atomic layer deposition at 200 °C. The temperature rises of ITO transistors were measured using SThM, consisting of a commercial module from Anasys® Instruments integrated with the MFP-3D AFM from Asylum Research. All measurements were performed in passive mode, where the sample was heated by electrical biasing. The SThM tip recorded temperature-dependent changes in its electrical resistance, and produce a voltage signal ($V_{SThM}$), which was then converted to a temperature rise ($\Delta T$) through a calibration process listed in Supporting information **Section S3**. The SThM scans were conducted in contact mode at a setpoint of 0.5 V and a scan rate of 0.8 Hz. Measurements were performed at room temperature (~20 °C) in air under 20-30% humidity. The thermal probe used in this work (PR-EX-GLA-5, from Anasys® Instruments) is made of a thin Pd layer on SiN.

**Finite-Element Modeling.** The finite-element electro-thermo-mechanical simulations were conducted using COMSOL Multiphysics®. Steady-state simulations were conducted with the bottom of the Si substrate set as thermal ground. The ITO channel was modeled with a uniform sheet resistance across its width. We used nominal thermal conductivity ($k$) and thermal boundary conductance ($G$) values as listed in Supporting Information **Table S1**. Some $G$ values not available in literature were approximated by values for pairs of similar and/or better-studied materials. Sensitivity analysis was carried out by varying $k$ and $G$ values to understand their impact. In addition, the strain distribution in ITO transistors during operation was assessed using the COMSOL thermal expansion model, with mechanical properties listed in Supporting Information **Table S2**. The initial strain induced by Ni contact was assessed by wafer-level Ni stress characterization, with further details provided in Supporting Information **Section S5**.

## SUPPORTING INFORMATION

Supporting information includes fabrication process flow (Section S1), additional electrical breakdown characteristics (Section S2), scanning thermal microscopy details (Section S3), finite-element electro-thermo-mechanical modeling (Section S4), wafer-scale Ni stress characterization (Section S5), devices on 30 nm $HfO_2$ dielectric (Section S6), and breakdown behavior of ALD-grown ITO devices (Section S7).

## AUTHOR DECLARATIONS

### Author Contributions

H.S. and Y.L. designed the experiments, supported by E.P. H.S. and Y.L. fabricated the samples and performed electrical characterization with support from T.P., J.K., and S.W. H.S. performed SThM measurements and conducted finite-element simulations with input from S.F.-W., Ç.K., and E.P. H.S. wrote the initial manuscript, with input from Y.L. and E.P. All authors discussed the results and edited the manuscript.

### Conflict of Interest

The authors have no conflicts to disclose.

### Acknowledgments

Several authors were supported by PRISM and SUPREME, JUMP 2.0 centers sponsored by the Semiconductor Research Corporation (SRC) and DARPA. This work was also supported in part by the National Science Foundation under Grant No. 2345655 in the Partnerships for Innovation (PFI) program. The authors also acknowledge partial support from the Stanford Nonvolatile Memory Technology Research Initiative (NMTRI) Affiliates program. Part of this work was performed at Stanford Nanofabrication Facility (SNF) and Stanford Nano Shared Facilities (SNSF), supported by the National Science Foundation under award ECCS-2026822.

## DATA AVAILABILITY

The data that support the findings of this study are available from the corresponding author upon reasonable request.

16

# SUPPORTING INFORMATION

# High-field Breakdown and Thermal Characterization of Indium Tin Oxide Transistors

Haotian Su[1,†], Yuan-Mau Lee[2,†], Tara Peña[1], Sydney Fultz-Waters[2], Jimin Kang[1], Çağıl Köroğlu[1], Sumaiya Wahid[1], Christina J. Newcomb[3], Young Suh Song[1], H.-S. Philip Wong[1], Shan X. Wang[1,2], and Eric Pop[1,2,4,5,*]

[1]*Department of Electrical Engineering, Stanford University, Stanford, CA 94305, USA*

[2]*Department of Materials Science and Engineering, Stanford University, Stanford, CA 94305, USA*

[3]*Stanford Nano Shared Facilities, Stanford University, Stanford, CA 94305, USA*

[4]*Department of Applied Physics, Stanford University, Stanford, CA 94305, USA*

[5]*Precourt Institute for Energy, Stanford University, CA 94305, USA*

[†]*Authors contributed equally.* [*]*Contact:* epop@stanford.edu

## S1. Fabrication Process Flow

The fabrication process flow is illustrated in **Figure S1** and additional details are provided in the **Methods** section (main text). The deposition rates for all steps were calibrated using X-ray reflectivity. For scanning thermal microscopy (SThM) measurements, the devices were capped with a 6 nm $Al_2O_3$ layer deposited via plasma-enhanced atomic layer deposition at 200 °C.

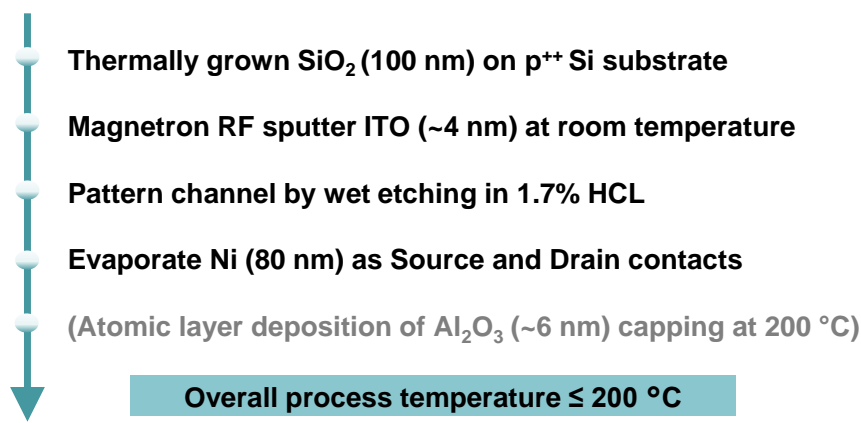

**Thermally grown $SiO_2$ (100 nm) on p$^{++}$ Si substrate**

**Magnetron RF sputter ITO (~4 nm) at room temperature**

**Pattern channel by wet etching in 1.7% HCL**

**Evaporate Ni (80 nm) as Source and Drain contacts**

**(Atomic layer deposition of $Al_2O_3$ (~6 nm) capping at 200 °C)**

**Overall process temperature ≤ 200 °C**

**Figure S1**. Fabrication process flow for ITO transistors, highlighting key steps. The entire process was conducted at temperatures ≤ 200 °C.

## S2. Additional Electrical Breakdown Characteristics

We conducted electrical breakdown measurements on ITO transistors under various biasing conditions; the transfer curves before and after breakdown are shown in **Figures S2a-e**. The transfer curves after breakdown under different breakdown biasing consistently exhibit a significant reduction in on-state $I_D$ and loss of gate control. Gate leakage current ($I_G$) was also monitored during all measurements and showed negligible change under breakdown conditions, as illustrated in **Figure S2f**. This suggests that the breakdown of ITO transistors is not due to gate dielectric breakdown.

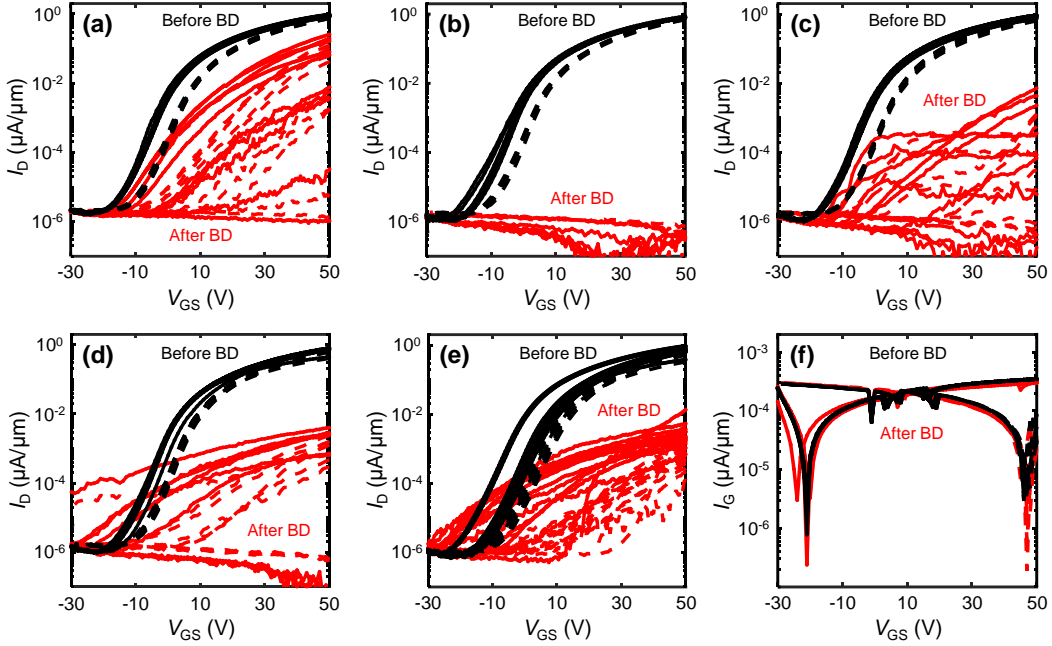

**Figure S2**. Transfer curves ($I_D$ vs. $V_{GS}$, measured at $V_{DS} = 0.1$ V) of ITO transistors measured before (black) and after breakdown (red) under different breakdown biasing conditions: **(a)** $V_{GS} = -15$ V and $V_{DS} = 40$ V; **(b)** $V_{GS} = 0$ V and $V_{DS} = 40$ V; **(c)** $V_{GS} = 15$ V and $V_{DS} = 40$ V; **(d)** $V_{GS} = 60$ V and $V_{DS} = 40$ V; **(e)** $V_{GS} = 75$ V and $V_{DS} = 25$ V. Each figure displays ~10 measured devices. Transfer curves after breakdown demonstrate a significant reduction in on-state $I_D$ and loss of gate control. All devices have channel length of 1.6 μm. Solid/dashed lines are forward/backward $V_{GS}$ sweeps. **(f)** Gate leakage current ($I_G$) vs. $V_{GS}$ before (black) and after (red) breakdown under $V_{GS} = 45$ V and $V_{DS} = 40$ V, showing similar trends for both cases, which rules out gate dielectric breakdown as the failure mechanism. (Note the different vertical axis scale, compared to the other figure panels.)

## S3. Scanning Thermal Microscopy (SThM) Details

We measured the temperature rise of ITO transistors using scanning thermal microscopy (SThM). The SThM system employs a thermo-resistive probe connected to a Wheatstone bridge, a DC voltage source, and an amplifier specifically designed to minimize electrical spikes that could damage the probe.[1-3] The ITO transistors were capped with 6 nm $Al_2O_3$ to prevent electrical shorting to the SThM probe. (Additional details are provided in the **Methods** section of the main text.) Although the SThM measures the surface temperature of this $Al_2O_3$ rather than the underlying ITO channel, prior studies have shown that such thin capping layers have a negligible effect on the measurement,[1,4] because the surface temperature of the $Al_2O_3$ layer closely matches the temperature of the ITO channel. To confirm this, we also conducted simulations with and without the capping layer, further verifying that the capping layer does not alter the temperature rise.

The SThM measurements produce a voltage signal ($V_{SThM}$) that correlates with the sample surface temperature. To convert $V_{SThM}$ to the corresponding temperature rise ($\Delta T$), we previously calibrated the same SThM probes using samples with a set of Ti/Pd heaters of varying widths, as described by S. Deshmukh *et al*.[3] The conversion factor ($F$, with the unit of mV/K) is determined using the known temperature coefficient of resistance (TCR) of the metal lines. These heaters were capped with an $Al_2O_3$ layer similar to that on our ITO transistors, to account for similar thermal (boundary) resistance at the probe-sample interface. While the conversion factor ($F$) can vary between probes, these variations are more obvious only for measurements on sub-200 nm features.[3] For the micron-scale ITO transistors in this study, $F$ remains consistent across probes. Based on our calibration, we adopted a conversion factor $F = 7.0 \pm 0.5$ mV/K for this study.

**S4. Finite-element Electro-Thermo-Mechanical Modeling**

We estimate the temperature rise ($\Delta T$, above room temperature) of our ITO transistors using three-dimensional finite-element electro-thermo-mechanical modeling, through COMSOL Multiphysics®.[5] To simplify the simulation, the ITO channel was modeled with a uniform sheet resistance across their width. Additionally, contact resistance is negligible in terms of heat generation because the ITO transistors in this study are sufficiently long[6] ($L \approx 1.5$ to $1.7$ μm). The thermal conductivity ($k$) of the materials and thermal boundary conductance ($G$) for various material interfaces used in our simulations are listed in **Table S1**.

**Table S1**. Nominal thermal properties used in simulations, including thermal conductivity ($k$) and thermal boundary conductance ($G$) values, all near room temperature. Some specific $G$ values not available in the literature were approximated by $G$ values for pairs of similar and/or better-studied materials.[5] The $G$ of ITO-SiO$_2$ and ITO-HfO$_2$ were ultimately estimated to be $35 \pm 12$ MWm$^{-2}$K$^{-1}$ and $51 \pm 14$ MWm$^{-2}$K$^{-1}$, respectively (see main text **Figure 5** and Supporting Information **Section S6**).

| Material | $k$ (Wm$^{-1}$K$^{-1}$) | Material Interface | $G$ (MWm$^{-2}$K$^{-1}$) |
|---|---|---|---|
| p$^{++}$ Si[1,7,8] | 95 | Si-SiO$_2$ | 500 |
| SiO$_2$[9,10] | 1.4 | ITO-SiO$_2$ | 50 |
| ITO[11-15] | 2 | ITO-Ni | 150 |
| Ni[16,17] | 40 | Ni-Al$_2$O$_3$ | 150 |
| Al$_2$O$_3$[18,19] | 1.5 | ITO-Al$_2$O$_3$ | 50 |
| HfO$_2$[8,20-22] | 1.1 | Si-HfO$_2$ | 283 |
|  |  | ITO-HfO$_2$ | 50 |

The assumption of 2 Wm$^{-1}$K$^{-1}$ for the thermal conductivity of ITO, as well as the values for Al$_2$O$_3$ and Ni, may not be entirely precise. To assess their impact, we carried out sensitivity analysis in **Figure 5b** of the main text, varying the thermal conductivity of ITO (1–10 Wm$^{-1}$K$^{-1}$), Al$_2$O$_3$ (1.2–3 Wm$^{-1}$K$^{-1}$), and Ni (15–60 Wm$^{-1}$K$^{-1}$). The results show that these variations have a negligible effect on the device temperature rise during operation, within the typical power inputs used here, suggesting that our conclusions remain robust despite uncertainties in material properties. This is not surprising, because for our long-channel devices heat dissipation is "vertical" into the substrate, not "lateral" along the ITO and into the Ni contacts.

We estimate the electronic contribution to the ITO thermal conductivity with the Wiedemann-Franz law, $k_e \approx 0.03$ Wm$^{-1}$K$^{-1}$ at ~180 °C, indicating that phonon transport dominates in our ultrathin ITO. Given this, we expect the total thermal conductivity of ITO to fall within the 1–10 Wm$^{-1}$K$^{-1}$ range, consistent with prior literature. The impact of Ni thermal conductivity is minimal, because our transistors are sufficiently long for heat dissipation to be dominated by the substrate rather than the contacts.[23] Similarly, even a $\pm 20\%$ change in the thermal boundary conductance ($G$) of Si-SiO$_2$ and Si-HfO$_2$ causes less than $\pm 1\%$ change in the calculated ITO channel temperature rise ($\Delta T$), because the dominant thermal resistance of our transistors is due to their underlying SiO$_2$ or HfO$_2$. Sensitivity analysis of other $G$ values is shown in the main text **Figure 5c**.

To investigate the strain distribution in ITO transistors during operation, we also conducted mechanical simulations by including the COMSOL thermal expansion module. The bottom of the Si substrate was set as the fixed boundary and linear elastic material properties were assumed for all constituent materials. The nominal mechanical properties, including the coefficient of thermal expansion ($\alpha$), Young's modulus ($E$), and Poisson's ratio ($v$), are listed in **Table S2**. Additionally, the initial stress induced by the Ni layer was accounted for in the simulations, with further details provided in Supporting Information **Section S5**.

**Table S2**. Nominal mechanical properties used in simulations, including the coefficient of thermal expansion ($\alpha$), Young's modulus ($E$), and Poisson's ratio ($v$) for constituent materials in the ITO transistor.

| Material | $\alpha$ (K$^{-1}$) | $E$ (GPa) | $v$ |
|---|---|---|---|
| SiO$_2$[24,25] | $5.6 \times 10^{-7}$ | 70 | 0.17 |
| HfO$_2$[26,27] | $6 \times 10^{-6}$ | 250 | 0.25 |
| ITO[28-31] | $8 \times 10^{-6}$ | 250 | 0.33 |
| Ni[32-35] | $1.3 \times 10^{-5}$ | 200 | 0.30 |
| Si[24,25] | $2.6 \times 10^{-6}$ | 170 | 0.28 |

## S5. Wafer-scale Ni Stress Characterization

An 80 nm Ni layer was deposited via electron-beam evaporation onto a 350 µm thick (100) Si wafer with native oxide, under a base pressure of ~5 × 10$^{-8}$ Torr. The wafer curvature was measured both before and after metal deposition. Using the Stoney equation,[36] the stress in the Ni film was found to be ~75 MPa. The thin film force was also determined, defined as $F = \sigma t$, where $\sigma$ is the film stress and $t$ is the film thickness. The measured Ni thin film force, summarized in **Table S3**, align well with the values reported in the literature.[37,38]

**Table S3**. Extracted stress and thin film force for the 80 nm Ni layer deposited as metal contacts.

| Metal, nm | Stress (MPa) | Thin film force (N/m) |
|---|---|---|
| Ni, 80 | 75 | 6 |

## S6. Devices on 30 nm HfO$_2$ Dielectric

We also fabricated sputtered ITO transistors on 30 nm HfO$_2$ back-gate dielectric on Si (p$^{++}$) back-gate substrates, as shown in **Figure S3a**, using the same fabrication flow detailed in Supporting Information **Section S1**. The 30 nm HfO$_2$ layer was deposited by plasma-enhanced atomic layer deposition at 200 °C, and its thickness was calibrated by ellipsometry. Measured electrical transfer curves and output characteristics are shown in **Figures S3b-c**. The Al$_2$O$_3$ capping layer was found to induce a negative shift in the threshold voltage, consistent with our previous study.[6] Combing SThM and finite-element simulations, we determined the TBC between the ITO channel and the HfO$_2$ dielectric to be 51 ± 14 MWm$^{-2}$K$^{-1}$, as shown in **Figure S3d**. **Figure S3e** shows the relationship between $\Delta T_{max}$ and input power ($P$), demonstrating good agreement between temperatures measured by SThM and simulated temperature at the breakdown power ($P_{BD}$). The breakdown temperature ($T_{BD}$) of ITO transistors on 30 nm HfO$_2$/Si substrate is found to be ~272-340 °C. This $T_{BD}$ was nearly double that of devices on 100 nm SiO$_2$/Si substrates. Mechanical simulations of strain distribution along the ITO channel **Figure S3f** revealed compressive strain under the contacts and tensile strain in the channel region under initial conditions. Under breakdown conditions, the peak compressive strain was observed near the channel edges. Our results show that ITO transistors on $HfO_2$ substrates exhibit significantly higher breakdown power, enabled by enhanced heat dissipation and closer thermal expansion matching between ITO and $HfO_2$.

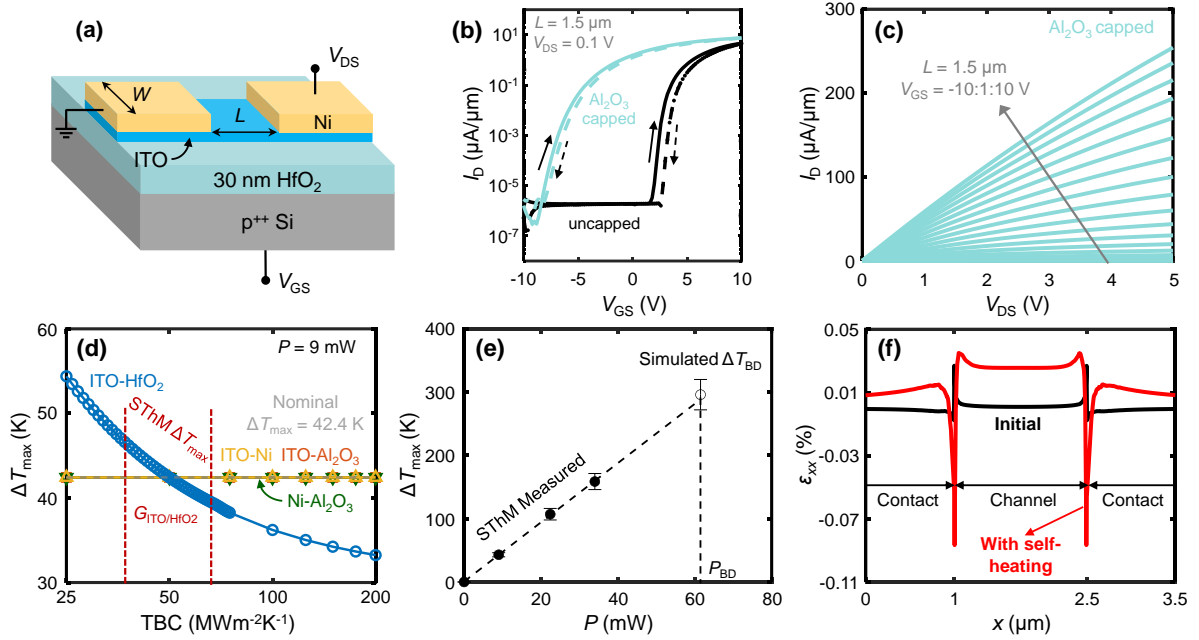

**Figure S3.** **(a)** Schematic of back-gated transistor, with 4 nm ITO channel on 30 nm $HfO_2$/Si ($p^{++}$) substrate, and 80 nm Ni source and drain contacts. **(b)** Measured transfer curves of uncapped (black) and $Al_2O_3$-capped (light blue) devices, showing negative shift in threshold voltage, similar to our previous work.[6] **(c)** Measured $I_D$ vs. $V_{DS}$ curve of $Al_2O_3$-capped devices, showing linear behavior. **(d)** Sensitivity analysis of $\Delta T_{max}$ with respect to TBC at material interfaces, showing minimal dependence on TBC at ITO-Ni, ITO-$Al_2O_3$, and Ni-$Al_2O_3$ interfaces. However, $\Delta T_{max}$ varies with the TBC of the ITO-$HfO_2$ interface, estimated as $51 \pm 14$ MWm$^{-2}$K$^{-1}$ by comparing to the SThM measurements. **(e)** Dependence of $\Delta T_{max}$ on input power, $P$. Filled symbols mark SThM-measured temperature, hollow symbol is a simulation of $\Delta T_{max}$ at $P_{BD}$. Dashed line is a linear fit, to highlight the trend. **(f)** Initial strain distribution ($\varepsilon_{xx}$) along the channel direction, showing tensile strain in the ITO channel and compressive strain under contacts (black curve). At device breakdown with self-heating, the compressive strain peaks at the channel/contact edge (red curve).

We note that the estimated TBCs for ITO-$SiO_2$ (~35 MWm$^{-2}$K$^{-1}$) and ITO-$HfO_2$ (~51 MWm$^{-2}$K$^{-1}$) are near the lower bound of typical TBCs for material interfaces,[39] but consistent with relatively low TBCs at the $In_2O_3$-$HfO_2$ interface[40] and those of other wide band gap material interfaces.[41] This can be due to phonon density of states mismatch between the materials and/or to material microstructure near the interface (as well as bonding strength), which could depend on deposition conditions. In addition, if the current flow in the ITO thickness is non-uniform, the effective TBC may appear lower by ~10% (e.g. in the limit of current flowing entirely at the *top* ITO surface).

### S7. Breakdown Behavior of ALD-grown ITO Devices

While this study mainly focused on the electro-thermal behavior of *sputtered* ITO devices, we also fabricated and characterized 3 nm thick ITO transistors grown by atomic layer deposition (ALD). **Figure S4a** illustrates the schematic of such back-gated devices on $SiO_2$, patterned with varying channel lengths into transfer length method (TLM) structures. Electrical measurements up to breakdown did not affect neighboring devices, and the breakdown voltage and current of 1-2 μm length channels were comparable to those of sputtered ITO transistors (**Figure 3** of main text).

Similar to the behavior of sputtered ITO transistors (main text), we also observe mechanical cracks from the drain side into the channel of ALD-grown ITO transistors (**Figure S4b-c**). While direct comparisons between sputtered and ALD-grown ITO transistors are limited due to differences in ITO thickness and electrical properties (e.g., threshold voltage), these findings suggest that similar high-field breakdown mechanisms occur in ALD-grown ITO transistors.

Given that ITO consists of 9:1 $In_2O_3$:$SnO_2$ by weight, it may be reasonable to expect that other In-dominated thin films transistors (e.g. $In_2O_3$, indium tungsten oxide, indium zinc oxide) have similar breakdown behaviors, although this topic warrants further research. On the other hand, thermal boundary conductance (TBC) is known to depend on factors such as material composition, interfacial bonding strength, and deposition conditions, making it difficult to predict TBC values for different amorphous oxide semiconductor interfaces with their gate insulators and contacts. Calculating TBC between amorphous materials from first principles also remains challenging (due to the lack of well-defined atomic structures and disordered interfacial bonding), thus we expect that experimental measurements will be needed for other (future) oxide semiconductor interfaces.

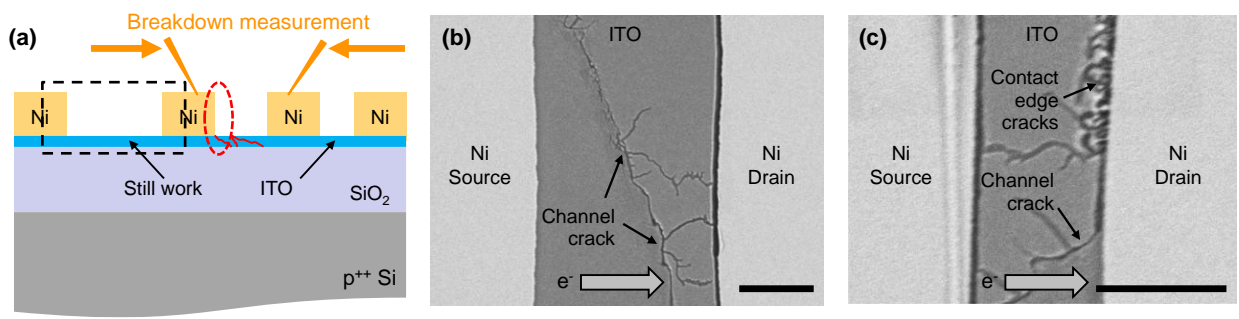

**Figure S4.** **(a)** Schematic of back-gated transistors with 3 nm *ALD-grown* ITO channel on 100 nm $SiO_2$/Si ($p^{++}$) substrate, here patterned into TLM structures. **(b)** Scanning electron microscopy (SEM) image of a device after breakdown showing channel cracks, and **(c)** another device showing additional cracks near the Ni drain contact. Scale bars in (b-c) are all 1 μm. Block arrows in (b, c) show the direction of electron flow. The observed breakdown mechanism is very similar to that of *sputtered* ITO devices with comparable channel lengths (**Figure 2** of the main text).

## SUPPORTING REFERENCES